\documentclass[preprint,showpacs,showkeys,preprintnumbers,amsmath,amssymb]{revtex4} 
\usepackage{bm}
\usepackage{epsfig,amssymb}
\newcommand{\sech}{\mbox{\rm sech}}

\begin{document}
\title{Solitary wave solutions as a signature of the instability in
  the discrete nonlinear  Schr\"odinger equation}
\author{Edward Ar\'evalo}
\affiliation{Technische Universit\"at Darmstadt, Institut f\"ur
  Theorie elektromagnetischer Felder, TEMF, Schlo{\ss}gartenstr. 8
  D-64289 Darmstadt, Germany}
\date{\today}
\begin{abstract}
The effect of instability on the
propagation of solitary waves along one-dimensional discrete
  nonlinear Schr\"odinger equation with cubic nonlinearity
is revisited. A self-contained quasicontinuum
approximation is developed to derive closed-form expressions for
  small-amplitude solitary waves.  The notion
  that the existence of nonlinear solitary waves in discrete systems 
is a   signature of  the modulation instability is used. With the help
of this notion we conjecture that instability
  effects on moving solitons can be qualitative estimated from the
  analytical solutions. Results from numerical simulations are
  presented to support this conjecture.
\end{abstract}
\maketitle


\section{introduction}
The discrete nonlinear  Schr\"odinger equation (DNLSE) is one
of the most investigated systems in dynamical nonlinear lattices. In
particular the DNLSE with cubic nonlinearity is an ubiquitous
dynamical-lattice system which has been extensively studied because
its direct physical applications, such as Bose-Einstein condensates
(BECs) in deep optical lattices or optical beams in waveguide arrays,
among others. In particular, the transport
properties of neutral atoms in BEC arrays has gained interest in the
last few years \cite{Fano07} due, in principle, to the possible technological
applications as matter-wave
interferometry \cite{Schumm05} or quantum information processing 
\cite{kay06,kay06A}.

Of course, solitary waves are promising candidates for
the coherent matter wave transport in BEC arrays. 
Notice that, in the strict sense, solitons (defined as
solitary waves that
maintain their shape and amplitude owing to a self-stabilization
against diffraction through a nonlinear interaction) do not exist in the
DNLSE. It is because the DNLSE is a nonintegrable system. However, approximate
moving solitary-wave solutions can be analytically calculated. One of
the main characteristic of these waves is that they emit radiation
when moving, and eventually become pinned by the lattice
\cite{oxtoby07}.  This is a  radiative deceleration which is extremely
slow \cite{oxtoby07}, and therefore do not play a role in the present
analysis.

In the following we shall use a weak definition of soliton to
refer to solitary wave solutions following from approximate
integration of the DNLSE.

We note that diverse analytical approximations have been 
developed in the last two decades to obtain moving solitary wave solutions of
the DNLSE for bright solitons (the theory of the perturbed Ablowitz-Ladik equation
\cite{yu86,YuSK93,ch93,Aceves96,gomez2004}, iteration methods
\cite{ablowitz2002}, semi-discrete approximation \cite{Peyrard04},
variational approximation \cite{Aceves96,BEC01}, or an explicit
perturbation theory 
\cite{ablowitz2002,oxtoby07,bifurcation05}) as well as for dark solitons
(quasicontinuum approximation \cite{kivshar94}, or
by construction \cite{Johansson99,bernardo05}). Of
course, the DNLSE in the continuum approximation leads to 
the standard nonlinear Schrodinger equation (NLSE). Notice, however, that the
soliton solutions of the NLSE do not move along the lattice, i.e. they
behave as breathers in the DNLSE system.

In the absence of dissipation and
inhomogeneities \cite{BEC01,BEC02,BEC03,firstband1,firstband2} and
taking into account that radiative deceleration is very slow \cite{oxtoby07}
the main effect
which reduces the life duration of these solitary waves is the instability of
the system. The instability effect in moving solitons manifests itself as a
distortion of exponential-like nature of the envelope during
propagation \cite{bifurcation05}.

It is  a common notion in discrete systems
that solitons exist or can be excited in the
region of parameters where modulation
instability (MI) of planewaves appears
\cite{Kivshar92,Kivshar92A,YuSK93,Konotop2002}. So, solitons by
themselves can be considered as a signature of the MI. Moreover, since
the MI is a permanent feature of the discrete system, i.e. it does not
disappear after the appearance of solitons, it is reasonable to
{\it conjecture} that the continuous deformation of moving solitons and
breathers after formation is a MI effect.


The usual analytical procedure for
studying MI effect on lattices is performing modulation stability 
analysis of planewaves 
\cite{Kivshar92,Kivshar92A,YuSK93,Konotop2002,discRep08,bernardo05}.
From this analysis usually qualitative conclusions are derived for
solitons. Recently the stability of narrow discrete modes in the DNLSE
was studied for the case $J\,U<0$ in Eq. (\ref{dnls1}) \cite{Pelinovsky05}.

Since solitons can be considered a signature of the MI effect in
discrete systems \cite{Kivshar92,Kivshar92A,YuSK93,Konotop2002}, it is 
natural to conjecture that approximate analytical forms of the soliton
solutions may contain already qualitative information of the MI (e.g. strength
and parameter region of existence). Here, the MI strength refers to the
grade of how during the time evolution the MI effect distorts the
initial analytical soliton solution.

In the present study we are interested in studying anew the the stability
of moving solitons in the DNLSE  with the help of approximate
analytical soliton solutions. In order to find these analytical
soliton solutions we use  a {\it self-contained} quasicontinuum
approximation (SCQCA). Our aim is to show that these approximate
analytical soliton  solutions  can be used to derive qualitative
conclusions of the soliton stability in the range of parameters where
the solutions are valid.  
In this regard we show that the amplitudes of these soliton solutions
define an analytical upper boundary for a ``self-defocusing'' instability not only
for bright but also for the dark solitons. Here the term
``self-defocusing'' instability refers to a decaying amplitude in moving
solitons which also undergo width broadening.
Numerical results supporting the analysis are presented. 
The standard modulation stability analysis is also revisited for
comparison issues.

We want to comment that it may sound superfluous to propose another
method for calculating solitary waves in the DNLSE since other
powerful methods as, e.g. those propose in Refs. 
\cite{oxtoby07,bifurcation05}, have already
captured in fine detail interesting aspects of this
equation. However, in contrast to those methods, the SCQCA can be 
straightforwardly extended to two \cite{edward09} and three spatial
dimensions, where no other analytical methods for moving solitons
have been proposed so far. Therefore,
understanding the effect of the modulation instability on the
solutions presented here is important in order to proceed to higher
dimensions.

\section{The SCQCA}
\subsection{the DNLSE}
The DNLSE with cubic nonlinearity, which is the basic model for 
the one-dimensional BEC arrays in the tight binding limit
\cite{BEC01,BEC02,BEC03,firstband2}, reads
\begin{equation}\label{dnls1}
i\partial_t \psi_n(t) +J (\psi_{n-1}(t)+\psi_{n+1}(t))
-U\vert \psi_n(t)\vert^2\psi_n(t)=0,
\end{equation}
where $\psi_n$ is a complex amplitude of the BEC mean field at the site
$n$, $J$ is proportional to the tunneling rate
\cite{BEC01,BEC02,BEC03,firstband1,firstband2} and  
the nonlinear coefficient $U$, known also as the interaction
strength. We note that Eq. (\ref{dnls1}) can be
written in a dimensionless form. However, since we are
interested in the sign effect of the nonlinear coefficient $U$ we keep
both parameters of  Eq. (\ref{dnls1}) in our analysis.

Equation (\ref{dnls1}) is a simple model which
reflects generic features of one-dimensional BEC arrays
with homogeneous scattering length. The effects of dissipation and
inhomogeneities are neglected
\cite{BEC01,BEC02,BEC03,firstband1,firstband2}. Expressions for 
$J$ and $U$ are known and depend, among others, on the
lattice spacing, the depth of the optical lattice, the $s$-wave scattering 
length, and the mass and number  of the atoms
\cite{BEC01,BEC02,BEC03,Fano07,firstband2}. 
These parameters can be manipulated in
experiments to obtain different desirable configurations of the system. 
In particular the interaction strength $U$ can be tuned 
through the $s$-wave scattering length, from positive
values (repulsive interactions) to negative values (attractive
interactions) by using  either magnetic or laser fields.

\subsection{Soliton solutions}

In the following we outline
the SCQCA for the DNLSE in the spirit of
Ref. \cite{Neuper94,Neuper94B,Neuper94A,edwardEPL08} to obtain approximate
analytical soliton 
solutions. As mentioned above, the DNLSE is a nonintegrable system, i.e. not
exact soliton solutions exist. Of course, the solitary solutions
derived below are only approximations, since SCQCA is only an
approximate integration of the DNLSE. We note, to the
best of our knowledge, that this SCQCA
\cite{Neuper94,Neuper94B,Neuper94A,edwardEPL08} has not been applied 
to the DNLSE. The importance of this method resides in the fact that
it can be used   without resorting to the knowledge of other
nonlinear partial differential equations. Therefore it is
self-contained.

In order to proceed with the SCQCA we consider a travelling wave
ansatz for an envelope complex function reading as
\begin{equation}\label{ansatz}
\psi_n(t)=\sum_{m=1}^{\infty}\chi_m(z)\exp({i m\theta}),
\end{equation}
where $z=n-v_0 v_k\,t$ and $\theta=k\, n-\epsilon_0 E_k\,t+\delta$. Here, 
$k$ is the quasimomentum, $v_k$ is a velocity, $E_k$ is the particle energy,
$\delta$ is a phase, and both $v_0$ and $\epsilon_0$ are
constants.

By using the full Taylor expansion of the
function  $(\psi_{n\pm 1} \rightarrow\exp(\pm\partial_n)\psi_{n})$,
Eq. (\ref{dnls1}) is transformed
into an operator equation,
\begin{equation}\label{dnls2}
i\partial_t \psi_n(t) +2 J \cos(\partial_n)\psi_{n}(t)
-W_n(t)=0,
\end{equation}
where $W_n(t)=U \vert \psi_n(t)\vert^2\psi_n(t)$.
Inserting the ansatz Eq. (\ref{ansatz}) in Eq. (\ref{dnls2}) and 
Fourier transforming the resultant equation we get
\begin{equation}\label{qca1}
\tilde{W}_m(q)=a_m(q)\tilde{\chi}_m(q),
\end{equation}
where
\begin{equation}\label{dnls3}
a_m(q)=2 J \cos(k m+q)+m \epsilon_0 E_k+q v_0 v_k
\end{equation}
Here tilde marks the Fourier transformed function, e.g.
\begin{equation}
\tilde{W}_m(q)=\frac{1}{\sqrt{2\pi}}\int\limits_{-\infty}^{\infty}dz e^{i q z} W_m(z),
\end{equation}
where $W_m$ is an abbreviation which collect all products of envelope
functions that belong to the same harmonic $e^{i m\theta}$, i.e.
\begin{equation}
\sum\limits_m W_m(z)e^{i m\theta}=W_n\left(\sum\limits_{l}\chi_l(z)e^{i
  l\theta}\right). 
\end{equation}

The SCQCA consist in  a formal solution of Eq. (\ref{qca1}) for
$\tilde{W}_m(q)$ and  in an expansion of the fraction $a_m(q)$
(Eq. (\ref{dnls3})) for small $q$ \cite{Neuper94,Neuper94B}, 
i.e. $a_m(q)\simeq \sum_n a_{nm}q^n$.  If we consider the expansion
of the fraction  $a_m(q)$ up to second
order in $q$, forcing the first order term of the expansion to be
zero by setting $v_0=1$, and transforming back to the position space,
we find a second-order  differential equation for the first
harmonic $\chi_1$, namely
\begin{equation}\label{elliptic1}
a_{01}\chi_1-a_{21} \partial_z^2\chi_1-3U
\vert\chi_1\vert^{2}\chi_1=0.
\end{equation}
Integrating by parts Eq. (\ref{elliptic1})  and
integrating again leads to a solution for the envelope:
\begin{eqnarray}\label{bright1}
\psi^B_n(z)=2 \sqrt{\frac{J \cos(k) (\epsilon_0-1)}{-3 U}}\sech
\left(\sqrt{2}\sqrt{(\epsilon_0-1)}z\right)e^{i \theta},
\end{eqnarray}
for $J\, U \cos(k)<0$  and $\epsilon_0>1$ (bright soliton),
\begin{eqnarray}\label{dark1}
\psi^D_n(z)=\sqrt{2}\sqrt{\frac{J\cos(k) (\epsilon_0-1)}{-3 U}}\tanh
\left(\sqrt{1-\epsilon_0}z\right)e^{i \theta},
\end{eqnarray}
for $J\,U \cos(k)>0$  and $\epsilon_0<1$ (dark soliton).
From the SCQCA we obtain also that $E_k=-2 J \cos(k)$ and $v_k=2 J
\sin(k)$. Notice that the range of high soliton velocity $v_k$ (rapid
solitons) corresponds to the values $|k|\simeq \pi/2$.

Here it is important to remark that Eqs. (\ref{bright1}) and/or
(\ref{dark1}) satisfy the so called staggering
transformation  $\psi_n\rightarrow (-1)^n \psi_n^{\star}$ ($\star$
stands for complex conjugation)
\cite{Johansson99}, which has been widely used in the literature to
consider the cases where the nonlinear coefficient $U$ changes the
sign. This transformation corresponds in
Eqs. (\ref{bright1}) and/or 
(\ref{dark1}) to the case $k\rightarrow \pi-k$. The solutions that
emerge after the transformation are similar to Eqs.  (\ref{bright1})
and/or (\ref{dark1}) but with the sign in front of $U$ changed. The
conditions for the bright and dark regions are also interchanged, i.e.  
$J\,U \cos(k)\rightarrow -J\,U \cos(k)$. Notice, however, that with
this transformation it is not possible to change from the bright region
to the dark region  or vice versa [equivalently, it is not possible to
  transform  Eq. (\ref{bright1}) into Eq. (\ref{dark1}) or vice
  versa]. So, in the present case, it is not appropriate to
infer conclusions from one region based on the information of the
other region by simply using this transformation.

It is worth mentioning that the  bright soliton solution,
Eq. (\ref{bright1}), is similar to that following from the  perturbed
Ablowitz-Ladik equation \cite{YuSK93,yu86,ch93,Aceves96,gomez2004}
only when $\cos(k)=1$ in the  amplitude of Eq. (\ref{bright1}). On the
other hand, a dependence of the amplitude on the quasimomentum $k$  was
obtained for bright solitons in  Ref. \cite{oxtoby07}, but only for
the case bright case ($J\,U\,\cos(k)<0$) in Eq. (\ref{dnls1}). In the case of
dark solitons, this dependence was obtained in Ref.
\cite{kivshar94}. So far, the  dependence of the soliton amplitudes on
the function $\cos(k)$ has not been analyzed in relation with their
stability.

It is also good to keep in mind that a single dark soliton, Eq. (\ref{dark1}), 
corresponds to the case where the sites of the BEC
array are equally filled except in the region where the hole excitation is
moving. However, one can construct pulse solitary waves in the dark
region by subtracting two identical dark solitons separated 
some distance $d>0$ (see Figs. \ref{fig4} and \ref{fig5}), i.e.
\begin{equation}\label{darkpulse1a}
u_n=\psi^D_{n+d/2}(t)-\psi^D_{n-d/2}(t).
\end{equation}
In the  following we shall call $u_n$ the dark pulse.
Notice that the width of $u_n$  depends not only on $k$ but also on $d$.
By choosing suitable values for $d$ and $\epsilon_0$ the form of dark
pulses $u_n$ can become very
similar (but not equal) to the form of bright solitons.

\section{Instability effect}

In this section we  first present a conjecture about how to estimate
the instability effect on the moving solitons. Afterwards, with the
help of numerical simulations we compare predictions from the
well-known  standard modulation stability analysis
\cite{Kivshar92,Kivshar92A,YuSK93,Konotop2002} 
 and from our conjecture. Special emphasis is done in the analysis of
 the dark 
region where the  standard modulation stability analysis fails to
predict analytically the instability of the system. We note that the instability of
staggered dark solitons was numerically studied in
Ref. \cite{Johansson99}.

\begin{figure}
\centerline{\epsfxsize=6.5truecm \epsffile{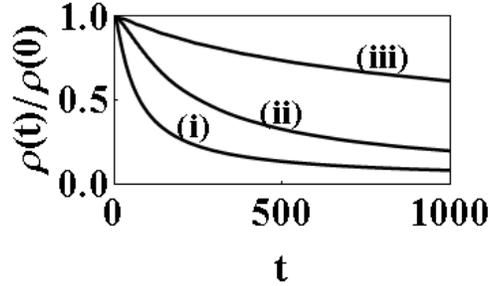}}
\caption{Normalized  probability-density maximum , $\rho(t)/\rho(0)$
  $[\rho(t)=\textrm{max}(\vert \psi^B_n(t)\vert^2])$,
  of the bright soliton 
  vs. the time $t$: (i) $k=0$ (or $\pi$), (ii) $k=2\pi/5$ (or $3\pi/5$),
  and (iii) $k=0.49\pi$ (or $0.51\pi$) with $J=1$ and  $\vert U\vert=1$.}
\label{fig0}
\end{figure}

\begin{figure}
\centerline{\epsfxsize=6.5truecm \epsffile{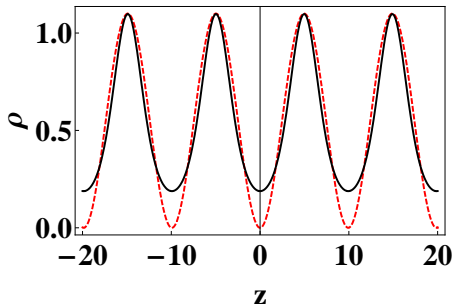}}
\caption{Probability density of an infinity train of bright
  solitons (solid line), i.e. $\rho=\vert \sum_j \psi^B_n(z-j L)\vert^2$, 
compared with a modulation plane-wave $\sin^2(Q)$ (dashed line). Here 
$Q=\sqrt{(\epsilon_0-1)}$ and $\epsilon_0=1.1$. The amplitudes are
  normalized to one.}
\label{fig1}
\end{figure}

\begin{figure}
\centerline{(a)\epsfxsize=3.0truecm \epsffile{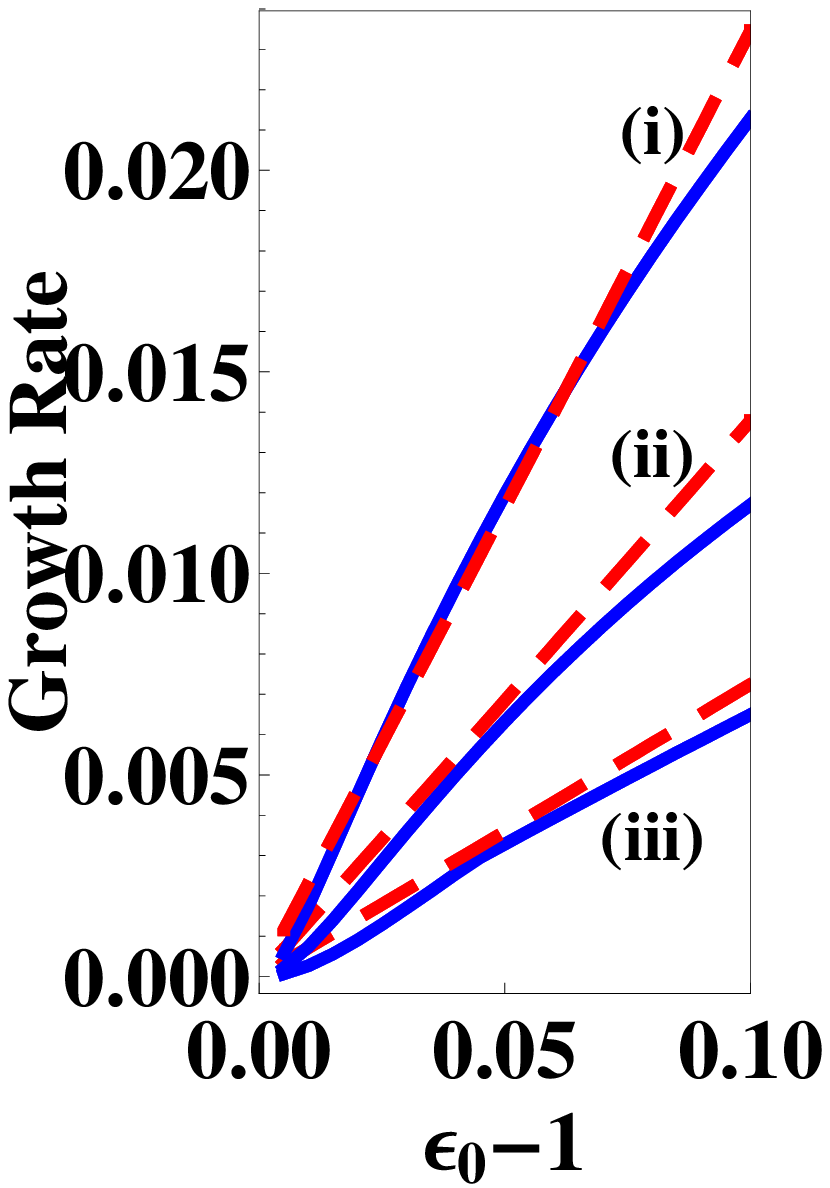}
(b)\epsfxsize=3.2truecm \epsffile{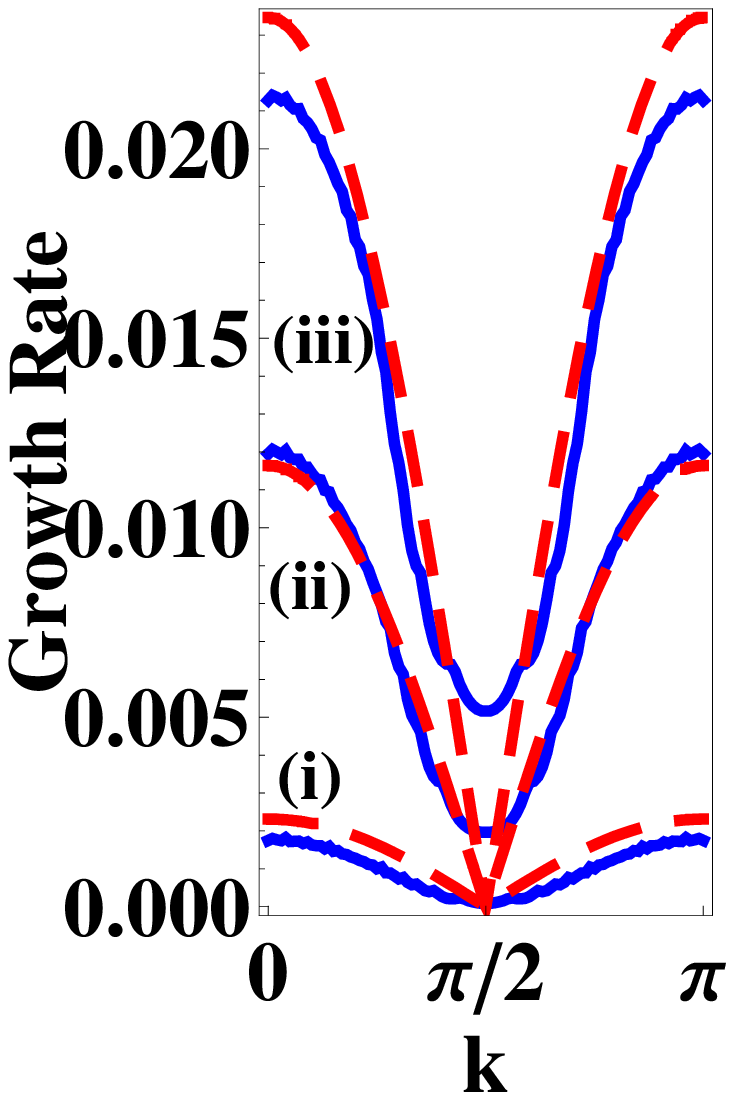}}
\caption{(Color online) Growth rate for bright solitons, $\psi^B_n$
  [Eq.(\ref{bright1})].  a: Versus $\epsilon_0-1$ for 
$(i)\, k=0\, (\textrm{or}\,\pi)$, $(ii)\, k=3\pi/10\,
(\textrm{or}\,7\pi/10)$, $(iii)\, k=2\pi/5\, (\textrm{or}\,3\pi/5)$. 
b: Versus $k$ for $(i)\, \epsilon_0=1.01$, $(ii)\, \epsilon_0=1.05$,
$(iii)\, \epsilon_0=1.1$. Numerical estimation
$\Gamma(T/2)$ with $T=10$ (solid line) and analytical estimation $\kappa_1 \Im(E_Q)$
(dashed line). $\kappa_1=0.4$ to fit numerical results.
}
\label{fig2}

\centerline{(a)\epsfxsize=3.0truecm \epsffile{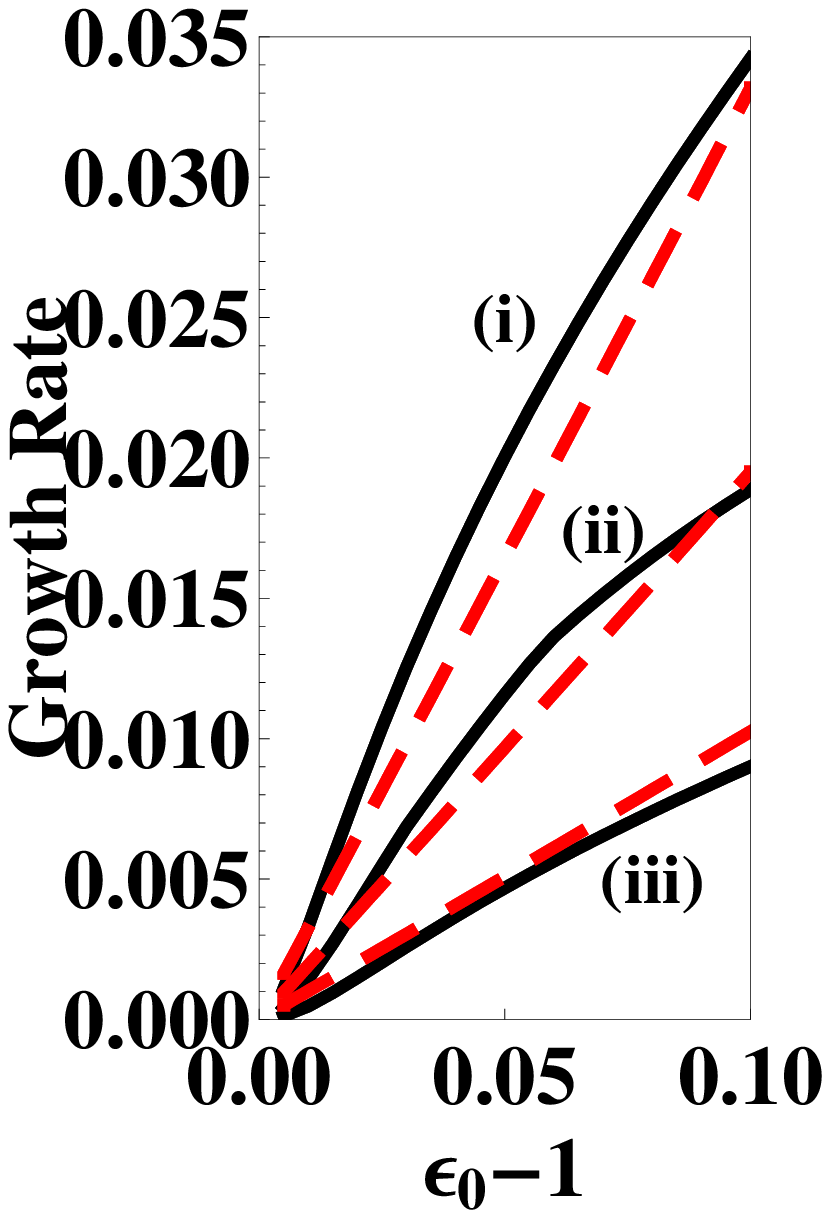}
(b)\epsfxsize=3.2truecm \epsffile{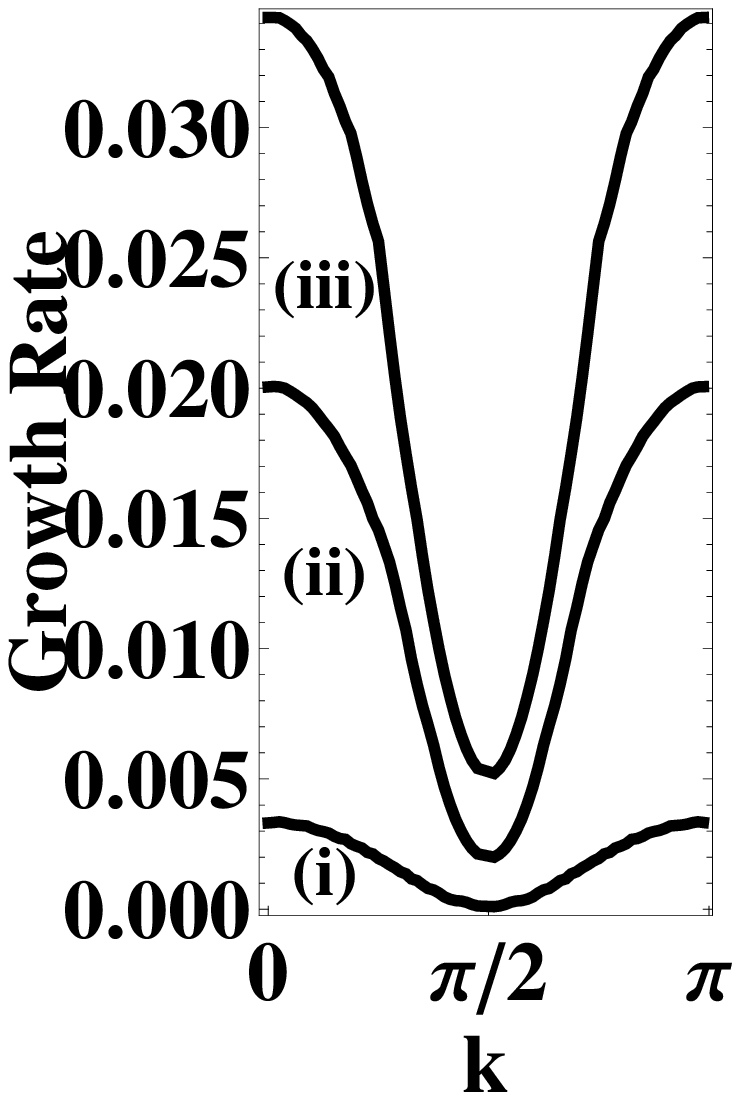}}
\caption{(Color online)  Numerical estimation of the Growth rate 
$\Gamma(T/2)$ with $T=10$ (solid line) in the dark region
for $u^{test}=i\psi^B_n$ [$i$ is the imaginary unit and $\psi^B_n$ is
given in Eq.(\ref{bright1})]. The same cases as in Fig. \ref{fig2} are
considered. In the panel (a) the analytical estimation for
bright  solitons $\kappa_2 \Im(E_Q)$ (dashed lines) is also plotted
with $\kappa_2=\sqrt{2}\kappa_1$ (see caption Fig. \ref{fig2}).}  
\label{fig3}
\end{figure}

Here we conjecture that the amplitudes of the approximate soliton solutions,
Eqs. (\ref{bright1}) and   (\ref{dark1}), contain 
qualitative information of the MI strength regardless of the soliton
form. Since  $\epsilon_0$ and $k$ are the free parameters of the system,
they govern the MI 
strength on the solitons. So, from a simple inspection of the amplitudes in 
Eqs. (\ref{bright1}) and
(\ref{dark1}) we observe that the simplest common factor in both solutions
containing $\epsilon_0$ and $k$ regardless of the signs is 
\begin{equation}\label{eta1}
\eta=\sqrt{|\cos(k) (\epsilon_0-1)|}.
\end{equation}
The absolute value in Eq. (\ref{eta1}) is used for convenience. Since
$\eta$ is proportional to the soliton amplitudes, we suppose
that the MI strength regardless of the region [bright ($J\,
U\,\cos(k)>0$) or dark ($ J\, U\, \cos(k)<0$)] is also proportional 
to $\eta$.

Notice that the magnitude of $\eta$ reduces as 
$|k|\rightarrow \pi/2$ vanishing
exactly at $|k|=\pi/2$ (maximum soliton velocity). So, here we can
conjecture that  MI for moving solitons
is present in the whole first Brillouin zone
except perhaps for $|k|=\pi/2$, i.e.  $|k|\in
[0,\pi/2)\cup(\pi/2,\pi]$. Moreover, since $\eta$
increases as $|\cos(k)|\rightarrow 1$, we can conjecture also that
the MI strength for moving solitons may  increase and reach its 
maximum at $|k|= 0,\pi$ (breather solutions) and may become weak as
$|\epsilon_0-1|\rightarrow 0$ (wide solutions). On the other hand, 
the MI effect on moving solitons can be expected regardless of the $J\, U\, \cos(k)$
sign, since both approximate bright and dark
soliton solutions have been derived.

The effect of MI on the small-amplitude soliton shapes [Eqs. (\ref{bright1}) and
(\ref{dark1})]  is of ``self-defocusing'' nature, i.e. a 
broadening of the soliton width accompanied by exponential-like decay
of the amplitude can be observed. As an example, in Fig. \ref{fig0} it is shown
the evolution of the normalized  probability-density maximum of bright
solitons for different
values of the quasimomentum $k$. This figure shows that the MI
strength can be characterized by studying the growth rate of the soliton
amplitudes. Since
in the present theory only defocusing MI effect is observed, the growth
rates for moving solitons are decaying \cite{bifurcation05}.

We note that solitary waves with higher
amplitudes than those of Eqs. (\ref{bright1}) and
(\ref{dark1}) can undergo other MI effects as, e.g., self-trapping effect
\cite{Aceves96,BEC01} and  strong oscillatory instabilities
\cite{Aceves96,bernardo05}. 


In order to check our conjectures above we
proceed to estimate a growth rate from 
the standard modulation stability analysis of plane waves in
one-dimensional arrays
\cite{Kivshar92,Kivshar92A,Konotop2002,discRep08}. Here, one looks at 
the  dispersion relation of a phonon wave whose amplitude and phase are 
perturbed by a modulation plane-wave. This yields  a
dispersion relation that for decaying small-amplitude waves read as
\cite{Kivshar92,Kivshar92A,discRep08}
\begin{eqnarray}\label{dispersion1}
&&\Omega_Q=2 J \sin(k)\sin(Q) -\sqrt{8 J}\times\nonumber\\
&&\sqrt{U \psi_0^2 \cos(k)\sin^2(Q/2)+2 J \cos^2(k)\sin^4(Q/2)},
\end{eqnarray}
where  $\Omega_Q$ and $Q$ are the frequency and wavenumber, respectively,
of the modulation plane-wave. $\psi_0$ is the amplitude of
the phonon wave. The growth rate $\Gamma$ of the modulation wave can be estimated 
from the imaginary part $\Im(\Omega_Q)$ of Eq. (\ref{dispersion1}), 
i.e. $\Gamma\sim\Im(\Omega_Q)$. 
Notice that $\Im(\Omega_Q)\ne 0$ if $J\,U \cos(k)<0$ (bright region) and 
$Q \in (0,2 \arcsin[\sqrt{-U\psi_0^2\sec(k)/(2 J)}])$.

At first glance, the comparison between a phonon wave modulated by a
plane-wave and the single soliton solutions in Eqs. (\ref{bright1}) and
(\ref{dark1}) may appear ``strange''. However, both problems are
strongly linked \cite{Kivshar92,Kivshar92A,YuSK93,Konotop2002}.
Notice that the modulated phonon wave in Eq. (\ref{dispersion1}) can be seen as
infinity train $\psi=\sum_j \psi^B_n(z-j L)$ of bright
solitons, where $L$ is a distance between the solitons. In
Fig. \ref{fig1} an example of this comparison is shown.
Similar construction can be done in the dark region
with the help of an infinity train of dark pulses [see
  Eq. (\ref{darkpulse1a})] to emulate a modulated phonon
wave.

A simple estimation of $\Im(\Omega_Q)$ for
bright solitons [Eq. (\ref{bright1})] can be done by 
assuming in Eq. (\ref{dispersion1}) that $\psi_0$ is
the soliton amplitude [$\psi_0\equiv
\sqrt{{4 J \cos(k) (\epsilon_0-1)}/(-3 U)}$]  
and $Q$ is proportional to the inverse width [$Q \equiv
  \sqrt{(\epsilon_0-1)}$]. We note that the approximate estimation of
$Q$ can be done by comparing the envelope shape of  a half-cycle
oscillation of the modulation wave with the shape of  the bright-soliton
envelope (see Fig. \ref{fig1}).

Notice that the case $\Im(\Omega_Q)=0$ (dark region: $U \cos(k)>0$) 
means that plane waves are stable. However, if one tries to extend this
conclusion for solitons, one would wrongly conclude that dark
solitons are stable. This wrong
notion would imply that MI effect is negligible in the dark region, which
contradicts the fact that approximate dark solitons
can be systematically derived. Moreover, instability for staggered
dark solitons have been numerically predicted \cite{Johansson99}.
Hence, we can conclude that the usual
modulation stability analysis, Eq. (\ref{dispersion1})
\cite{Kivshar92,Kivshar92A,YuSK93,Konotop2002}, is not appropriate for 
deriving conclusions on the stability of solitons in the dark region.

It has been argued that the staggering
transformation $(k\rightarrow \pi-k)$ in Eq. (\ref{dispersion1})  may
overcome this problem. However, as mentioned above, by using this
transformation the conditions for the bright and dark regions are
also automatically interchanged. In other words, with this
transformation one changes from one bright region to other
bright region. So,  in the present case the staggering
transformation in combination with the usual modulation stability
analysis cannot analytically describe MI effect on moving solitons in
the dark region.

The remaining question is: how the MI effect for moving solitons
can be analytically described in the  dark region?  In order to
investigate this question,  
in the following, we numerically estimate the growth rate associate
with the MI strength for both the bright and the dark regions.
Afterwards, we compare numerical results with estimations from the
standard modulation stability analysis and from the the term $\eta$ in 
Eq. (\ref{eta1}). We note that the analysis in the bright region is
presented only for completeness, and in order to compare  with
results obtained in the dark region.

From numerical simulations in Fig. \ref{fig0}
we can observe that the growth rate depends 
on the time, $\Gamma=\Gamma(t)$, in
a non-trivial form. In fact, the 
exponential-like nature of the MI
effect observed in the probability density,
$\rho_n=|\psi_n^B|^2$, can be described by a exponential
function of the form $\rho_n \sim \exp(-2\int_0^{t}\Gamma dt^{\prime})$.

The growth rate can be estimated in simulations by using the 
expression $\Gamma(t)=-d\rho_{max}(t)/(2 \rho_{max}(t) dt)$, where   
$\rho_{max}(t)=\int_{t-T/2}^{t+T/2} \textrm{max}(\rho_n(t))dt^{\prime}/T$ is
an average in time of the maximum of $\rho_n$.  This average is used to 
smooth out small
oscillations of $\rho_n$ due to the discreteness of the system and is
valid for $t\ge T/2$. Here $T$ is a small time scale, i.e. $2/T \gg
\Gamma(T/2)$. Notice that these small oscillations in time of
the solitons have been discussed in Ref. \cite{ablowitz2002}.

Since the analytical estimation of the growth rate 
[$\Gamma\sim\Im(\Omega_Q)$] can be
done only for the initial soliton form, we perform a comparison only for 
a short initial time scale, namely  $t=T/2$.

In Fig. \ref{fig2} we show the numerical estimations of the 
initial $\Gamma$ value, i.e.  $\Gamma(T/2)$, for the bright region . A
comparison with the analytical from the standard modulation analysis 
is also presented. We
observe that except for a constant factor ($\kappa_1=0.4$) the 
analytical approximation $\kappa_1\Im(E_Q)$ describes qualitatively 
well the behavior of the growth rate. 
In particular, in Fig. \ref{fig2}a it is possible to
observe that the strength of the MI effect reduces 
as the soliton width (amplitude) increases (decreases), i.e. 
$\epsilon_0-1\rightarrow 0$. 
In Fig. \ref{fig2}b estimations of the growth
rate for the first Brillouin zone are presented, showing that MI 
effects become weak as $|\cos(k)|\rightarrow 0$, but they do not
completely vanish as predicted by the theory. However, $\Gamma$ tends
to be negligible for very small amplitude solitons, i.e $\epsilon_0 \simeq
1$, $k\simeq \pi/2$. Notice that qualitative predictions from $\eta$
[Eq. (\ref{eta1})]  are identical to the  standard modulation analysis
in the present case.

Now we consider the case of MI effect on solitons in the dark region
($J\,U\,\cos(k)>0$), where Eq. (\ref{dispersion1}) does not provide any
information. However, since we have found approximate dark soliton
solutions, we should expect also MI effects here. Besides, since
$\eta$ [Eq. (\ref{eta1})] is equal in both regions, we can expect that
the behaviour MI strength on pulse solitons in the dark region is
similar to that in the bright region.

Notice that in order to obtain more information about the MI effect on
solitons, it is convenient to compare both the bright and dark regions. 
An {\it exact} quantitative comparison of a bright soliton [Eq.
(\ref{bright1})] with a dark pulse, $u_n$ [Eq. (\ref{darkpulse1a})], is 
not possible, since their envelopes cannot not be exactly equal. So, 
for the sake of comparison and before we characterize 
the dark pulses $u_n$, we define a sech
pulse in the  dark region as $u_n^{test}=i\psi_n^B$. Here  $i$ is
the imaginary unit and $\psi_n^B$ is given in Eq.(\ref{bright1}). This
pulse is identical to a bright soliton, except that the relation
$J\,U\,\cos(k)>0$ can be considered.

In Fig. \ref{fig3}a and \ref{fig3}b
we show numerical estimations of $\Gamma$ vs. $\epsilon_0-1$ and $k$,
respectively, for $u_n^{test}$. The same cases  as in Fig. \ref{fig2}
are considered here too. We observe that Fig.  \ref{fig3} is
qualitatively similar to Fig. \ref{fig2}. However, the $\Gamma$ values
in the dark region (Fig.  \ref{fig3}) have  a factor $\sqrt{2}$ larger
than in the bright region (Fig.  \ref{fig2}). At first glance, it  is
a surprising result, taking into account that plane waves in the dark
region are stable. However, it is straightforward to
observe that $\eta\sim A_B/2$ or $\eta\sim A_D/\sqrt{2}$,
where $A_B$ and $A_D$ are the bright [Eq. (\ref{bright1})]
and dark [Eq. (\ref{dark1})] amplitudes, respectively. It means that
for identical bright and dark pulses with equal amplitude
($A_B=A_D$), as in Figs. \ref{fig2} and \ref{fig3}, the
MI strength, represented by $\eta$ is $\sqrt{2}$ times higher in the
dark region than in the bright one.
It also means that in order to obtain in the dark region a quantitative 
similar MI result than in Fig. \ref{fig2} (bright region), 
one can define a sech pulse in the
dark region but with the amplitude and width of the dark soliton
[Eq. (\ref{dark1})], regardless of the soliton form. This assumption  
can be  straightforwardly corroborated by numerical simulations.

The results above show that the strength of the MI effect in the
bright and dark regions is different. And this difference cannot be
estimated with the usual standard modulation stability analysis
in combination with the staggering transformation.
Moreover, it shows that the approximate soliton
solutions, in fact, do contain information regarding their modulation
instability. To the best of
our knowledge, this type of analysis for moving solitons in the dark
region and in the whole first Brillouin zone has not been reported, so far.

\begin{figure}
\centerline{\epsfxsize=3.5truecm \epsffile{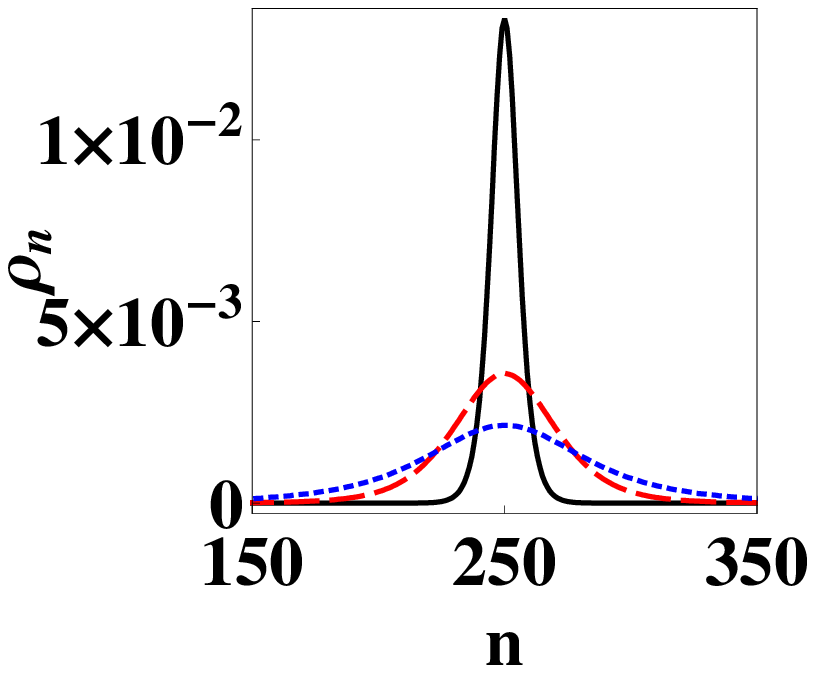}
\epsfxsize=3.5truecm \epsffile{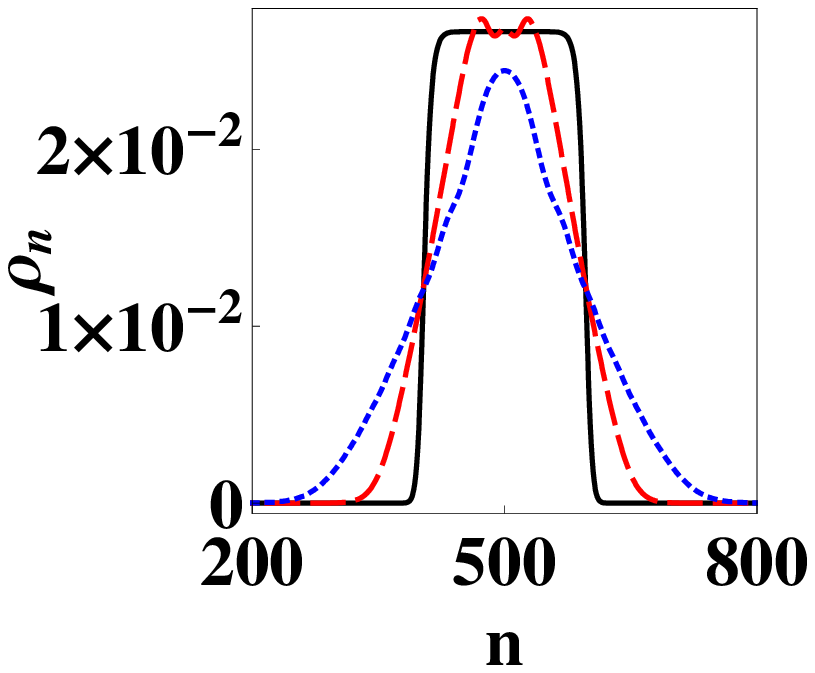}}
\caption{(Color online) Snapshots of the
probability density $\rho_n$ of a Bright (left panel)  and 
dark (right panel) breathers at different times [$t=0$ (solid line),
200 (dashed line), and 400 (dotted line)]. 
$J=1$, $U=-1$, $|\epsilon_0-1|=0.01$,
 $k=0$ (left panel), and $k=\pi$ (right panel).}
\label{fig4}
\end{figure}

\begin{figure}
\centerline{\epsfxsize=6.5truecm \epsffile{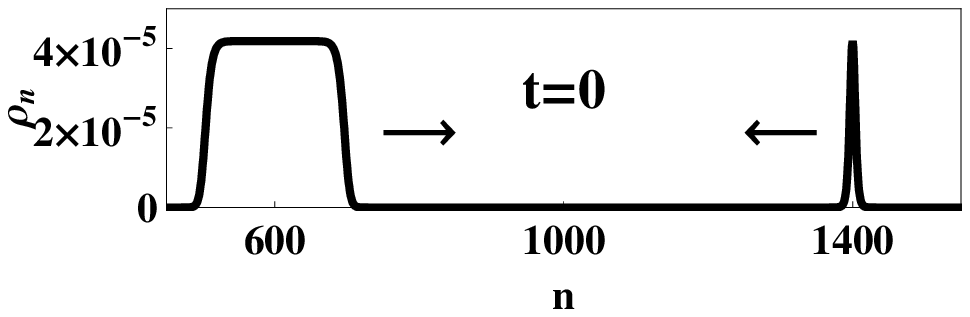}}
\centerline{\epsfxsize=6.5truecm \epsffile{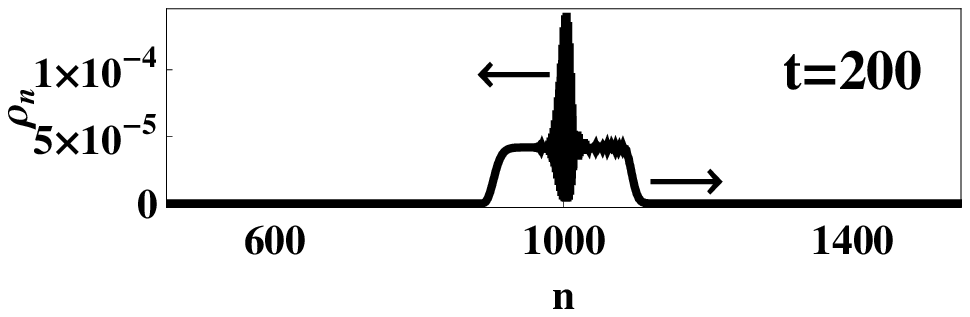}}
\centerline{\epsfxsize=6.5truecm \epsffile{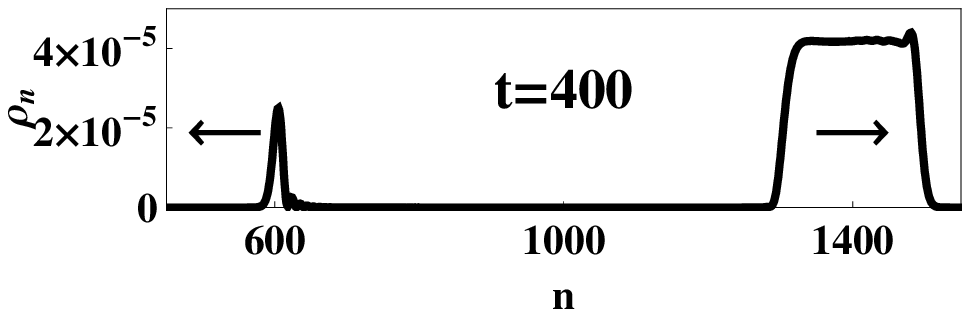}}
\caption{Snapshots of the probability density $\rho_n$
of a collision between a bright soliton [Eq. (\ref{bright1})]  and a dark 
pulse [see
Eqs. (\ref{darkpulse1a}) and (\ref{dark1})] at different times (top-bottom: 
$t=0,200$, and 400). Arrows indicate direction of motion: 
``$\rightarrow$'' for the dark pulse and ``$\leftarrow$'' 
for the bright soliton.
 $k=0.49 \pi$ (bright), $k=0.51\pi$ (dark),
$\epsilon_0=1.02$ (bright), $\epsilon_0=0.99$ (dark), $J=1$, and
 $U=-1$.}
\label{fig5}
\end{figure}

For completeness in Figs. \ref{fig4} and \ref{fig5} snapshots
of the probability-density evolution  of solitons
[Eq. (\ref{bright1})]  are dark pulses [Eq. (\ref{darkpulse1a})] are shown.
In particular, in Fig
\ref{fig4} two breather cases ($|\cos(k)|=1$) for  the bright (Fig. \ref{fig4}a)
and dark region (Fig. \ref{fig4}b) are shown. we
observe the usual broadening of the pulse and the decaying 
of the amplitude at different time scales. It is important
to remark
that in the case of the dark region (Fig. \ref{fig4}b) the collapse of
the breather starts from the lateral sides, and then moves to the center.
Notice that the breather maximum does not decay immediately but
when the collapse process reaches the center.

In Fig. \ref{fig5} an example of a collision between a 
high-velocity bright soliton and dark
pulse is shown. We observe that those high-velocity solitons
are stable against collisions and they are less prone to MI effect, as
expected from the analysis done for \ref{fig2}, and \ref{fig3}.

For very long time scales solitons in Fig. \ref{fig5} undergo
similar collapse as for the breather case in Fig. \ref{fig4}.
We note that the small amplitude radiation observed in
Fig. \ref{fig5} have been investigated with great detail in
Refs. \cite{oxtoby07}.

Since, the solitons from Eqs. (\ref{bright1}) and (\ref{dark1})
undergo only  ``self-defocusing'' instability, we can conclude that
these solutions pose an analytical upper boundary for
``defocusing-like'' solitary waves of the 
DNLSE. So, solitary waves with higher amplitudes can undergo other
instabilities as the oscillatory or the self-trapped
\cite{Aceves96,BEC01,bernardo05}.

Finally, we note that the soliton solutions in Eqs. (\ref{bright1}) and
(\ref{dark1}) vanish at $|k|=\pi/2$. This value, known as the 
``zero-dispersion'' point, has been examined with great
detail in Ref. \cite{bifurcation05}.

\section{Conclusions}

Motivated by the problem of coherent matter wave transport in
BEC arrays, within the first-band approximation,
we have studied anew the problem of modulation instability (MI) of
small-amplitude solitary waves in the discrete Schr\"odinger
equation (DNLSE).   For that
we have developed a self-contained quasicontinuum approximation
(SCQCA) for the DNLSE to
derive approximate analytical soliton solutions. We have used the well-known
notion that solitons can be considered as a  signature of the MI
to conjecture that analytical soliton solutions following from
the approximate integration of the system contain already
qualitative information of the MI strength on the solitons when
propagating. We have shown with the help of numerical simulations that this
conjecture describes qualitatively well the ``self-defocusing'' MI effect of
small-amplitude solitons in the dark region where the standard 
modulation stability analysis of planewaves and/or the staggering transformation
do not provide any information. Though planewaves are stable in the dark region, we
have shown that for identical pulses in the bright and dark region
the strength of the ``self-defocusing'' MI is higher in the dark region than
in the bright one. This fact was shown not only numerically but also
by following the conjecture posed above.

Since the soliton solutions derived here only present
``self-defocusing'' instability, their amplitudes can be
considered as analytical upper boundaries for this
instability.

Last but not the least, the analysis proposed here can be 
straightforwardly extended to higher spatial dimensions (work in
progress) where the motion of solitary waves and vortices can be
also observed and analyzed.

\end{document}